\documentclass[allclo]{FBSart}
\usepackage{amsfonts}
\usepackage{amssymb}
\usepackage{graphicx}
\usepackage{epsfig}

\newcommand{\beq}{\begin{equation}}
\newcommand{\eeq}{\end{equation}}

\title{Dynamical coupled-channels: the key to understanding 
resonances}
\author{B. Juli\'a-D\'\i az\thanks{\textit{Alternative address:} Excited Baryon Analysis Center, 
Thomas Jefferson National Accelerator
Facility, Newport News, Va. 22901}}
\institute{Departament de Estructura i Constituents de la Materia,\\
University of Barcelona,
Spain, 08028}

\runningauthor{B. Juli\'a-D\'\i az}
\runningtitle{Dynamical coupled-channels: the key to understanding resonances}
\begin{document}

\maketitle
\begin{abstract}
Recent developments on a dynamical coupled-channels model of hadronic and electromagnetic
production of nucleon resonances are summarized. 
\end{abstract}

\section{Introduction}

The analysis of nucleon resonance production and subsequent 
decay is possibly the best tool to study the 
most striking aspects of non-perturbative QCD. Following this 
fact a large multinational effort has been devoted during 
the last two decades to gather detailed information about 
resonance excitation both using hadronic and electromagnetic probes. 

In recent years most of the emphasis has fallen on the 
electroexcitation of resonances seen in meson electroproduction 
experiments,~\cite{lee-reviewa,lee-reviewb}, and a large amount 
of accurate data has been collected at Jefferson Laboratory (JLab), 
Mainz, Bonn, GRAAL, and Spring-8. The involved nature of the problem, 
which comprises excitation of resonances and multistep meson-baryon 
processes is nowadays believed to demand a dynamical coupled-channels 
approach. In this work we will consider the one developed 
recently, MSL, and which is explained in great detail in Ref.~\cite{msl}. 

The starting point of the MSL model is a set of Lagrangians 
describing the interactions between mesons (including the photon) 
($M$ =$\gamma$, $\pi, \eta$ , $\rho, \omega$, $\sigma, \dots$) and 
baryons ($B = N, \Delta, N^*, \dots$). By applying a unitary transformation 
method~\cite{sl-1}, an effective Hamiltonian, 
with an energy independent set of potentials, is then derived from 
the considered Lagrangian.

The resulting meson-baryon ($MB$) scattering amplitudes are, 
\begin{eqnarray}
 T_{\alpha,\beta}(E)  &=&  
 t_{\alpha,\beta}(E)
+ 
 t^R_{\alpha,\beta}(E) \,,
\label{eq:tmbmb}
\end{eqnarray}
where $\alpha, \beta = \gamma N, \pi N, \eta N, \pi\pi N$. The full
amplitudes, e.g. $T_{\pi N,\pi N}(E)$, $T_{\eta N,\pi N}(E)$, 
$T_{\pi N,\gamma  N}(E)$ can be directly used to, within 
the same framework, compute $\pi N$, $\pi N \to \eta N$ and 
$\gamma N \to \pi N$, $\gamma N \to \eta N$, scattering observables. 
The non-resonant amplitude $t_{\alpha,\beta}(E)$ in Eq.~(\ref{eq:tmbmb}) is
defined by the coupled-channels equations,
\begin{eqnarray}
t_{\alpha,\beta}(E)= V_{\alpha,\beta}(E)
+\sum_{\delta}
V_{\alpha,\delta}(E) \;
G_{\delta}(E)    \;
t_{\delta,\beta}(E)  \,
\label{eq:nr-tmbmb}
\end{eqnarray}
with
\begin{eqnarray}
V_{\alpha,\beta}(E)= v_{\alpha,\beta}
+Z^{(E)}_{\alpha,\beta}\,, 
\label{eq:veff-mbmb}
\end{eqnarray}
where $v_{\alpha,\beta}$ are the non-resonant $MB$ potentials and
$Z^{(E)}_{\alpha,\beta}$ is due to the one-particle-exchange between
unstable $\pi\Delta, \rho N, \sigma N$ states which are the resonant
components of the $\pi\pi N$ channel.

The second term in the right-hand-side of Eq.~(\ref{eq:tmbmb}) is the resonant 
term defined by
\begin{eqnarray} 
t^R_{\alpha,\beta}(E)= \sum_{N^*_i, N^*_j}
\bar{\Gamma}_{\alpha \rightarrow N^*_i}(E) [D(E)]_{i,j}
\bar{\Gamma}_{N^*_j \rightarrow \beta}(E) \,,
\label{eq:tmbmb-r} 
\end{eqnarray}
with
\begin{eqnarray}
[D^{-1}(E)]_{i,j} = (E - M^0_{N^*_i})\delta_{i,j} - 
 \sum_{\delta}\Gamma_{N^*_i\rightarrow \delta} G_{\delta}(E)
\bar{\Gamma}_{\delta \rightarrow N^*_j}(E) \,.
\label{eq:nstar-g}
\end{eqnarray}
where $M_{N^*}^0$ is the bare mass of the resonant state $N^*$.
The dressed vertex interactions in Eq.~(\ref{eq:tmbmb-r}) and
Eq.~(\ref{eq:nstar-g}) are (defining 
$\Gamma_{\alpha\rightarrow N^*}=\Gamma^\dagger_{N^* \rightarrow \alpha}$)
\begin{eqnarray}
\bar{\Gamma}_{\alpha \rightarrow N^*}(E)  &=&  
{ \Gamma_{\alpha \rightarrow N^*}} + \sum_{\delta}
t_{\alpha,\delta}(E) 
G_{\delta}(E)
\Gamma_{\delta \rightarrow N^*}\,, 
\label{eq:mb-nstar} \\
\bar{\Gamma}_{N^* \rightarrow \alpha}(E)
 &=&  \Gamma_{N^* \rightarrow \alpha} +
\sum_{\delta} \Gamma_{N^*\rightarrow \delta}
G_{\delta }(E)t_{\delta,\alpha}(E) \,. 
\label{eq:nstar-mb}
\end{eqnarray}

\section{Revisiting the electroexcitation of the $\Delta$(1232)}

\begin{figure}[t]
\vspace{10pt}
\begin{center}
\mbox{\epsfig{file=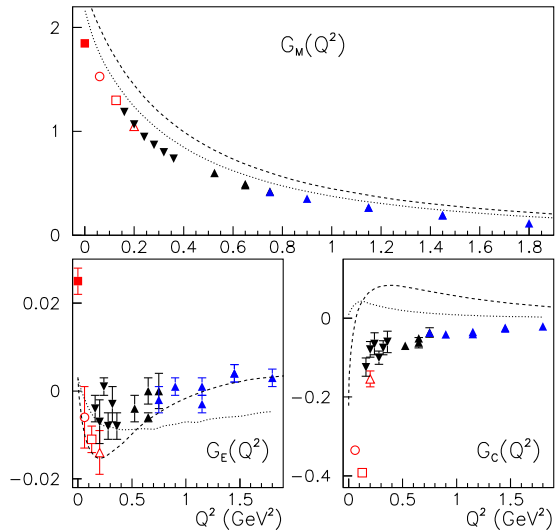, width=8cm}}
\end{center}
\caption{Bare form factors for the $\gamma N \to \Delta$ 
transition as a function of $Q^2$. The points have been 
obtained by performing individual fits for each $Q^2$ 
value to the corresponding pion electroproduction data 
given explicitly in Ref.~\cite{jlss07}. The dashed curves are 
from the front form quark model calculations of 
Ref.~\protect\cite{capstick-1}. The dotted curves are 
from the instant form quark model calculations of Ref.~\protect\cite{bruno-1}.
}
\label{fig:bff}
\end{figure}

To illustrate the procedure that is being followed in our current 
work we will first consider a simplified version of the model 
sketched above. We consider only two channels, $\gamma N$ and $\pi N$, 
and will concentrate on the $\Delta$ (1232) region. The full framework 
is described in great detail in Ref.~\cite{jlss07}. It is a revision 
of the Sato-Lee model~\cite{sl-1} with one important 
improvement: we extract the $N$-$\Delta$ form factors by fitting
all of the available pion electroproduction data at energies 
close to the $\Delta$ position.

The vertex $\bar{\Gamma}_{\gamma N \rightarrow N^*}(E)$ appearing in 
Eq.~(\ref{eq:mb-nstar}) can be written in terms of three form factors 
$\bar{G}_E(Q^2)$, $\bar{G}_M(Q^2)$ and $\bar{G}_C(Q^2)$ which are referred to 
as {\it dressed} form factors. Similarly the vertex, 
$\Gamma_{\gamma N \rightarrow N^*}(E)$, accepts a similar decomposition 
in terms of $G_E(Q^2)$, $G_M(Q^2)$ and $G_C(Q^2)$, which are called 
{\it bare} form factors. 

To extract the $N$-$\Delta$ dressed (and bare) form factors, we abandon 
the simple parameterization used in Ref.~\cite{sl-1} and perform $\chi^2$ 
fits to available experimental electroproduction data at each $Q^2$ 
by adjusting the values of the bare form factors. The resulting bare form 
factors are shown as symbols in Fig.~\ref{fig:bff}.

\section{Interpretation of the extracted form factors}

Like any reaction involving $composite$ systems, such 
as the atomic and nuclear reactions, the full amplitude 
describing the process has a non-resonant part and a 
resonant part, see Eq.~(\ref{eq:tmbmb}).

Qualitatively speaking, the non-resonant amplitude, $t$, 
is due to the {\it fast} process through some direct 
particle exchange mechanisms, and the resonant amplitude 
$t^R$ is due to the {\it time-delayed} process where the 
incoming particles lose their identities and form an 
unstable system which then decays into various final 
states. The unitarity condition ${\rm Im} T = T^\dagger T$ 
implies that $t$ and $t^R$ are not independent from 
each other. The close relation between the resonant and 
non-resonant amplitude is not specific to the formulation 
considered here, but is the consequence of a very general 
unitarity condition.

Thus the extracted dressed form factors $\bar{G}_M(Q^2)$, 
$\bar{G}_E(Q^2)$, $\bar{G}_C(Q^2)$ of the resonant amplitude 
can only be compared with the hadron structure calculations of 
current matrix element $<\Delta | j^\mu_{em}\cdot \epsilon_\mu |N> $
which contain meson loops. The bare form factors may in principle 
be compared to those obtained from phenomenological calculations 
which do not include meson clouds, e.g. quark models or specific 
lattice QCD simulations.

In Fig.~\ref{fig:bff} we compare our extracted bare form 
factors to two relativistic quark model studies, 
Refs.~\cite{capstick-1} and~\cite{bruno-1}. The comparison shows 
that both agree reasonably well for the magnetic and 
electric form factors but fail to capture the behavior of $G_C$. 
The next necessary step is to connect the current description 
to the extant lattice QCD simulations of the $\gamma N \to \Delta$ 
transition.  

\section{Full coupled-channels model (I): Meson-baryon interaction}
\label{sec:cal}

\begin{figure}[t]
\vspace{35pt}
\begin{center}
\mbox{\epsfig{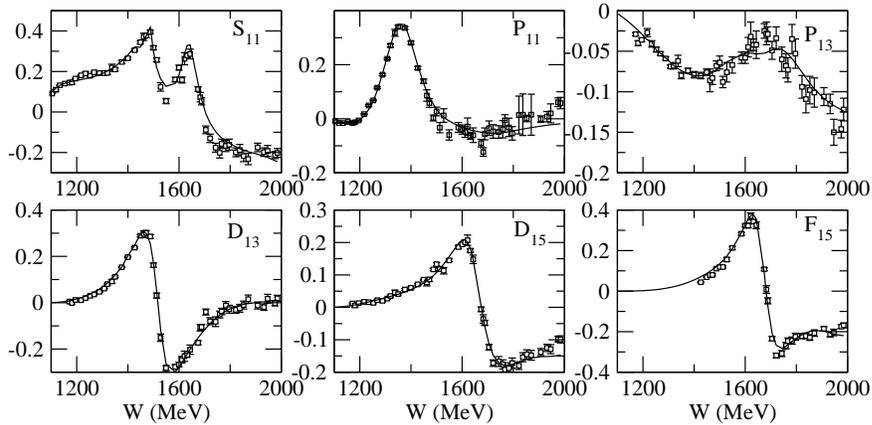}}
\end{center}
\caption{Real part of $T_{\pi N, \pi N}$ for some 
isospin $1/2$ partial waves compared to the SAID 
energy independent extraction.
\label{faa}}
\end{figure}

After the discussion in the previous section we go back to 
the full coupled-channels model which will be used 
to analyze the experimental electroproduction data at 
center of mass energies below 2 GeV. First we consider the 
meson-baryon interactions involving $\pi N, \eta N, (\pi \Delta, 
\sigma N, \rho N)$ and use the extensive database for 
$\pi N \to \pi N$ (and also the $\pi N \to \eta N$) to fix the 
non-resonant parameters entering in the phenomenological 
lagrangians. The parameter values used in this model 
are given in Ref.~\cite{jlms07}. Once the meson-baryon sector 
is fixed we will, in a first stage, leave it unchanged and 
produce a first description of the single meson photoproduction data.

With the non-resonant amplitudes generated from solving 
Eq.~(\ref{eq:nr-tmbmb}), the resonant amplitude $t^R_{MB,M'B'}$ 
Eq.~(\ref{eq:tmbmb-r}) then depends on the bare mass $M^0_{N^*}$ 
and the bare $N^*\rightarrow MB$ vertex functions, which 
are parametrized in Ref.~\cite{jlms07}.

In figure~\ref{faa} we depict the real part of $T_{\pi N, \pi N}$ 
compared to the energy independent extraction of the 
GWU group~\cite{said}, the agreement is quite good in almost 
all $S$, $P$, $D$, and $F$ waves. More importantly, the model 
describes successfully the differential cross section and 
target polarization asymmetries as shown in Ref.~\cite{jlms07}.

The fit to $\pi N$ elastic scattering cannot fully constrain the 
bare $N^* \rightarrow \pi \Delta, \rho N, \sigma N$ parameters. Thus 
the results for these unstable particle channels must be refined 
by fitting the $\pi N \rightarrow \pi\pi N$ data, this is currently 
being pursued~\cite{jklms07}.

\section{Full coupled-channels model (II): Photoproduction reactions}

With the hadronic parameters determined in the previous section 
we proceed to analyze the extensive database of $\pi$ photoproduction. 
Here the only parameters that need to be determined are the bare 
$\gamma N \to N^*$ vertex interactions of Eq.~(\ref{eq:mb-nstar}) similarly 
to what we did in the simplified model of Section 2.

The strategy is to start with the bare helicity amplitudes 
of resonances at the values given by the PDG~\cite{pdg}. Then, we 
allow small variations with respect to those values and also 
in a preliminary step also allow small variations of a selected 
set of non-resonant parameters. At this stage we can 
only present preliminary results which are at the present 
time being further improved and will be reported elsewhere. 

\begin{figure}[t]
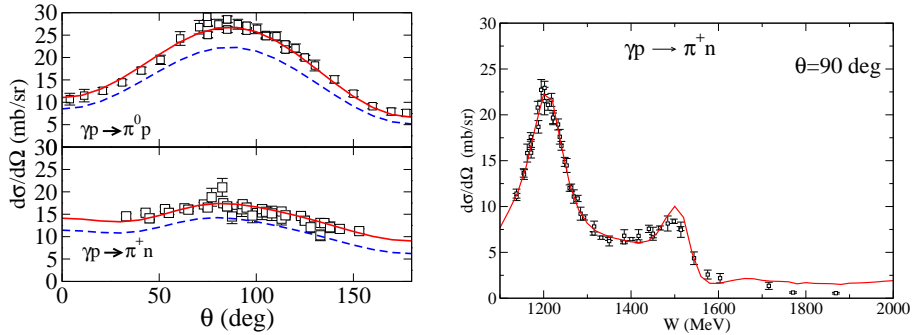

\vspace{25pt}
\begin{center}
\mbox{\epsfig{file=fig3a, width=54mm}}
\hspace{10pt}
\mbox{\epsfig{file=fig3b, width=60mm}}
\end{center}
\caption{(left) Differential cross section in the $\Delta$ region. The 
full line corresponds to the current full model, the dashed line 
only considers intermediate $\pi N$ intermediate states in the 
photoproduction process. (right) Differential cross section for 
a fixed angle.
\label{dcs_efb}}
\end{figure}

First, our main emphasis is set on understanding the region 
up to $1.6$ GeV but keeping a resonable description up to 2 GeV 
extending in that way previous works where only the $\Delta$ (1232) 
region was studied~\cite{sl-1,sl-2,jlss07}. In figure~\ref{dcs_efb} 
we depict angular distributions for both $\pi^+ n$ and $\pi^0 p$ 
photoproduction differential cross sections in the $\Delta(1232)$ region and 
also give an example of the prediction of the model for higher center of 
mass energies at a fixed angle. The effect of intermediate meson-baryon 
states different from $\pi N$ is also depict. The importance of 
multi-step processes is clear and confirms previous studies done 
in a similar framework.

\section{Future Developments}

The model described in detail in Refs.~\cite{msl,jlms07} has already 
been used to study $\pi N$ scattering and $\pi$ photoproduction reactions 
as presented in this contribution. In the near future we expect to extend 
these studies and perform a consistent study of meson-baryon scattering, 
single meson electro(photo) production~\cite{jlmss07} and two-meson photoproduction.

At the same time an important effort is being pursued to 
reliably extract meaningful resonance parameters from the 
coupled-channels formalism~\cite{suzukikun}.

\begin{acknowledge}
It is a pleasure to thank T.-S. H. Lee, A. Matsuyama, T. Sato and 
L. C. Smith who collaborated in most of what is presented here. I 
want to thank also the hospitality of the theory group at Jefferson Lab where 
part of this work was done. This work is partially supported by Grant 
No. FIS2005-03142 from MEC (Spain) and FEDER and European Hadron Physics 
Project RII3-CT-2004-506078. The computations were performed at NERSC (LBNL) 
and Barcelona Supercomputing Center (BSC/CNS) (Spain). 
\end{acknowledge}


\begin{thebibliography}{99}

\bibitem{lee-reviewa}
V. Burkert and T.-S. H. Lee, Int. J. of Mod. Phys. {\bf E13}, 1035 (2004).

\bibitem{lee-reviewb}
T.-S. H. Lee and L.C. Smith, J. Phys. G {\bf 34}, 1 (2007).

\bibitem{msl}
A. Matsuyama, T. Sato, T.-S. H. Lee, Phys. Rept. {\bf 439}, 193 (2007).


\bibitem{sl-1}
T.~Sato and T.-S.~H. Lee, Phys. Rev. C {\bf 54}, 2660 (1996);
T.~Sato and T.~S.~H.~Lee, Phys.\ Rev.\  C {\bf 63}, 055201 (2001).


\bibitem{jlss07}
B. Juli\'a-D\'\i az, T.-S. H. Lee, T.~Sato and L.~C.~Smith, 
Phys.\ Rev.\  C {\bf 75}, 015205 (2007).



\bibitem{capstick-1}
S. Capstick and B.D. Keister, 
Phys. Rev. {\bf 51}, 3598 (1995).

\bibitem{bruno-1}
B. Juli\'a-D\'\i az and D.~O.~Riska,
Nucl.\ Phys.\ A {\bf 757}, 441 (2005).

\bibitem{said}
 R.A. Arndt, I.I. Strakovsky, R.L. Workman,
 Phys. Rev. C {\bf 53}, 430 (1996);
 Int. J. Mod. Phys. {\bf A18}, 449 (2003); 
 CNS Data Analysis Center, GWU, {\bf http://gwdac.phys.gwu.edu}.

\bibitem{jlms07}
B. Juli\'a-D\'\i az, T.-S. H. Lee, A. Matsuyama, and T. Sato, 
 {\it arXiv:0704.1615}. To appear in Phys. Rev. C.



\bibitem{jklms07} B. Juli\'a-D\'\i az, H. Kamano, T.-S. H. Lee, 
A. Matsuyama, and T. Sato, in preparation.

\bibitem{pdg}
W.~M.~Yao {\it et al.}  [Particle Data Group], 
J.\ Phys.\ G {\bf 33}, 1 (2006).
         

\bibitem{sl-2}
T.~Sato and T.-S.~H. Lee, Phys. Rev. C {\bf 63}, 055201 (2001).

\bibitem{jlmss07}
B. Juli\'a-D\'\i az, T.-S. H. Lee, A. Matsuyama, T. Sato, L.C. Smith 
in preparation.


\bibitem{suzukikun} N. Suzuki, T. Sato and T.-S.H. Lee, in preparation.




\end{thebibliography}
\end{document}